\documentclass[pamm,a4paper,fleqn]{w-art}
\usepackage{times,cite,w-thm}
\usepackage[T1]{fontenc}
\usepackage[utf8]{inputenc}
\usepackage{upgreek}
\usepackage{amssymb}
\usepackage{color}
\newcommand\Rey{\mbox{\textit{Re}}}  
\newcommand\Ha{\mbox{\textit{Ha}}}  
\newcommand\Pran{\mbox{\textit{Pr}}} 
\newcommand\Ra{\mbox{\textit{Ra}}}
\newcommand\St{\mbox{\textit{St}}}
\usepackage{graphicx}
\begin{document}

\TitleLanguage[EN]
\title[The short title]{Simulation of magnetohydrodynamic flows of liquid metals with heat transfer or magnetic stirring}

\author{\firstname{Shashwat} \lastname{Bhattacharya}\inst{1}%
\footnote{Corresponding author: e-mail \ElectronicMail{shashwat.bhattacharya@tu-ilmenau.de},
     phone +49\,3677\,69\,2446},
\firstname{Seyed Loghman} \lastname{Sanjari}\inst{1,2},
\firstname{Dmitry} \lastname{Krasnov}\inst{1} and
\firstname{Thomas} \lastname{Boeck}\inst{1}
}
\address[\inst{1}]{\CountryCode[DE]Institute of Thermodynamics and Fluid Mechanics, TU Ilmenau, P.O. Box 100565, 98684 Ilmenau}
\address[\inst{2}]{\CountryCode[DE]CTWe GmbH, Kirchenstra{$\upbeta$}e, 91239 Henfenfeld}
\AbstractLanguage[EN]
\begin{abstract}
We discuss the effects of nonhomogeneous magnetic fields in liquid metal flows in two different configurations. In the first configuration, we briefly report the impact of fringing magnetic fields in a turbulent Rayleigh-B{\'e}nard convection setup, where it was shown that the global heat transport decreases with an increase of fringe-width. The convective motion in regions of strong magnetic fields is confined near the sidewalls. In the second configuration, we numerically study the effects of an oscillating magnetic obstacle with different frequencies of oscillation on liquid metal flow in a duct. The Reynolds number is low such that the  wake of the stationary magnetic obstacle is steady.
The transverse oscillation of the magnet  creates a sinusoidal time-dependent wake reminiscent of the vortex shedding behind solid obstacles.
We examine the behavior of the streamwise and spanwise components of the Lorentz forces as well as the work done by the magnets on the fluid.
The frequency of the oscillation of the streamwise component of Lorentz force is twice that of the spanwise component as in the case of lift and drag  on solid cylindrical obstacles.
The total drag force and the energy transferred from the magnets to the fluid show a non-monotonic dependence on the frequency of oscillation of the magnetic obstacle indicative of a resonant excitation of the sinusoidal vortex shedding.
\end{abstract}
\maketitle                   

\section{Introduction} \label{sec:Introduction}
Magnetohydrodynamic (MHD) flows, i.e., flows of electrically conducting fluids under the influence of magnetic fields, are frequently encountered in engineering and astrophysical applications. In such flows, the fluid is acted upon by the Lorentz force in addition to the force driving the flow.
Industrial and technological applications of such flows include heating, pumping, stirring, and levitation of liquid metals, cooling blankets in fusion reactors, and liquid-metal batteries.
In the context of astrophysics, magnetic fields strongly influence the flows in the sun and the stars and are responsible for the formation of sunspots and solar flares.

Magnetoconvection has been studied extensively in the past, but most of the studies focused on flows under the influence of a homogeneous magnetic field, which is an idealized approximation. However, in most engineering and astrophysical applications (such liquid metal batteries, cooling blankets in fusion reactors, electromagnetic stirring, solar spots, etc.) the magnetic fields are localized and thus vary in space~\cite{Davidson:ARFM1999,Davidson:book:MHD}.
Further, strong homogeneous fields in large regions of space can only be generated by magnets of large size which are difficult to design and very costly to build and operate~\cite{Barleon:KIT1996}. Thus, it is important to understand the impact of spatially varying magnetic fields on magnetohydrodynamic flows. Recently, Bhattacharya \textit{et al.}~\cite{Bhattacharya:JFM2023} studied the effects of spatially varying magnetic fields on MHD flows driven by buoyancy (magnetoconvection); these effects will be briefly summarized in this paper. There have been several studies on MHD duct flows with different configurations of spatially varying fields (see, for example, Sterl~\cite{Sterl:JFM1990} and Prinz \textit{et al.}~\cite{Prinz:PRF2016}); however, in this paper, we focus specifically on duct flows with a localized zone of applied magnetic field (henceforth referred to as \emph{magnetic obstacle}). Flows past stationary magnetic obstacles have been studied before~\cite{Cuevas:JFM2006,Votyakov:PRL2007,Evgeny:JFM2008,Kenjeres:IJHFF2011,Tympel:JFM2013}. Similarities and differences between stationary magnetic and solid obstacles has been discussed by Votyakov and Kassinos\cite{Votyakov:PF2009}. Unsteady wakes 
 were only found for fairly large Reynolds numbers where the flow develops small-scale turbulent eddies \cite{Kenjeres:IJHFF2011}. In order to realize  an unsteady flow past a  magnetic obstacle at a rather low Reynolds number it seems necessary to add an additional periodic motion of the magnet. We therefore consider the effects of \emph{oscillating} magnetic obstacles on MHD duct flow in the present paper, which can be interesting in the context of magnetic stirring. We also remark that oscillating solid obstacles have been studied previously but it appears that such studies are lacking for   magnetic obstacles so far.

The outline of the paper is as follows. In Sec.~\ref{sec:MathModel}, we discuss the mathematical model. Section~\ref{sec:Numerics} describes the numerical method used in our simulations. We discuss the results in Sec.~\ref{sec:Results} and conclude in Sec.~\ref{sec:Conclusions}.

\section{Mathematical model}\label{sec:MathModel}
In this section, we describe the setup and the mathematical formulation of our problems. The study will be conducted under the quasi-static approximation, in which the induced magnetic field is neglected as it is very small compared to the applied magnetic field. This approximation is fairly accurate for MHD flows of liquid metals~\cite{Davidson:book:MHD}. The governing equations of MHD flows are given by
\begin{eqnarray}
	\nabla \cdot \boldsymbol{u}&=&0, \label{eq:continuity} \\
	\frac{\partial \boldsymbol{u}}{\partial t}  + \boldsymbol{u}\cdot \nabla \boldsymbol{u} &=& -\nabla p + \nu \nabla^2 \boldsymbol{u}+ \boldsymbol{f},
	\label{eq:Momentum} \\
\end{eqnarray}
where $\boldsymbol{u}$ and $p$ are the velocity and pressure fields respectively, $\nu$ is the kinematic viscosity, and $\boldsymbol{f}$ is the total body force acting on the fluid.
For MHD duct flow, $\boldsymbol{f}$ is the specific Lorentz force (i.e. force per unit mass, henceforth denoted as $\boldsymbol{f}_L$) and is given by
\begin{equation}
	\boldsymbol{f} = \boldsymbol{f}_L =\frac{1}{\rho}(\boldsymbol{j} \times \boldsymbol{B}_0),
	\quad
	\boldsymbol{j} = \sigma(-\nabla \phi + \boldsymbol{u} \times \boldsymbol{B}_0),
	\quad
	\nabla^2 \phi = \nabla \cdot (\boldsymbol{u} \times \boldsymbol{B}_0).
	\label{eq:Lorentz}
\end{equation}
where $\boldsymbol{j}$ is the current density, $\boldsymbol{B}_0$ is the imposed magnetic field strength, $\sigma$ and $\rho$ are the electric conductivity and mean density of the fluid, respectively, and $\phi$ is the electric potential.

In magnetoconvection, $\boldsymbol{f}=\boldsymbol{f}_L+\boldsymbol{f}_b$, where $\boldsymbol{f}_b=\alpha g T \hat{z}$ is the buoyancy force, $\alpha$ is the thermal expansion coefficient, $g$ is the gravitational acceleration, and $T$ is the temperature field.
Magnetoconvection is additionally governed by the following thermal energy equation which describes the evolution of the temperature field $T$:
\begin{equation}
	\frac{\partial T}{\partial t}  + \boldsymbol{u}\cdot \nabla T = \kappa \nabla^2 T,
	\label{eq:Thermal}
\end{equation}
where $\kappa$ is the thermal diffusivity of the fluid.

MHD liquid-metal duct flows are governed by two nondimensional parameters: the Reynolds number $\Rey$, which is the ratio of the inertial force to the viscous force; and the Hartmann number $\Ha$, which is the ratio of the Lorentz force to the viscous force. Liquid-metal magnetoconvection is governed by three nondimensional parameters: the Rayleigh number $\Ra$, the ratio of the buoyancy force to the dissipative forces; the Prandtl number $\Pran$ -- the ratio of kinematic viscosity to the thermal diffusivity; and the Hartmann number $\Ha$. These quantities are given by
\begin{equation}
	\Rey = \frac{UL}{\nu},
	\quad
	\Ha = BL\sqrt{\frac{\sigma}{\rho \nu}},
	\quad
	\Ra = \frac{\alpha g \Delta L^3}{\nu \kappa},
	\quad
	\Pran = \frac{\nu}{\kappa},
	\label{eq:Nondimensional_numbers}
\end{equation}
where $U$, $L$, and $\Delta$ are the characteristic velocity, length, and temperature scales respectively. For magnetoconvection, we consider the Rayleigh-B{\'e}nard setup consisting of fluid enclosed between a cooler top plate and a warmer bottom plate (the temperature difference between the plates being $\Delta$), with the plates separated by a distance $H$. In this case, $H$ and $\Delta$ respectively are the characteristic length and temperature scales for $\Ha$ and $\Ra$.

As discussed in Sec.~\ref{sec:Introduction}, we describe the effects of spatially varying magnetic fields in liquid metal flow for two configurations - (i) thermal convection in a box, and (ii) pressure-driven duct flow. In the first configuration, we consider a horizontally extended convection box of size $l_x \times l_y \times H = 16 \times 32 \times 1$ which is influenced by magnetic fields generated by two semi-infinite permanent magnets. The north pole of one magnet faces the bottom of the convection cell and the south pole of the second magnet faces the top of the convection cell. These magnets extend from $-\infty$ to $\infty$ in the $x$-direction, $l_y/2$ to $\infty$ in the $y$-direction, from near the top wall to $\infty$ in the positive $z$-direction, and from near the bottom wall to $- \infty$ in the negative $z$ direction. For a detailed description of the setup, the readers are referred to Bhattacharya \textit{et. al}~\cite{Bhattacharya:JFM2023}.
In this configuration, the lateral component of the magnetic field ($B_x$) vanishes, and the longitudinal and vertical components respectively are logarithmic and inverse-tangent functions of the spatial coordinates $y$ and $z$ and the gap $\delta$ between the magnetic poles and the horizontal walls.
The magnetic field distribution is such that its strength is negligible for $0<y \lesssim l_y/2$, increases steeply at $y \sim l_y/2$, and saturates close to its maximum value at $y \gtrsim l_y/2$.
When $\delta$ is increased keeping other parameters same, the total magnetic flux through the convection cell remains the same, but the gradient of the magnetic field at $y\sim l_y/2$ decreases, thereby increasing the fringe-width of the magnetic field. The aim of the study was to determine the effects of fringe-width on the heat and momentum transport.

\begin{figure}
	\centerline{\includegraphics[scale = 0.32]{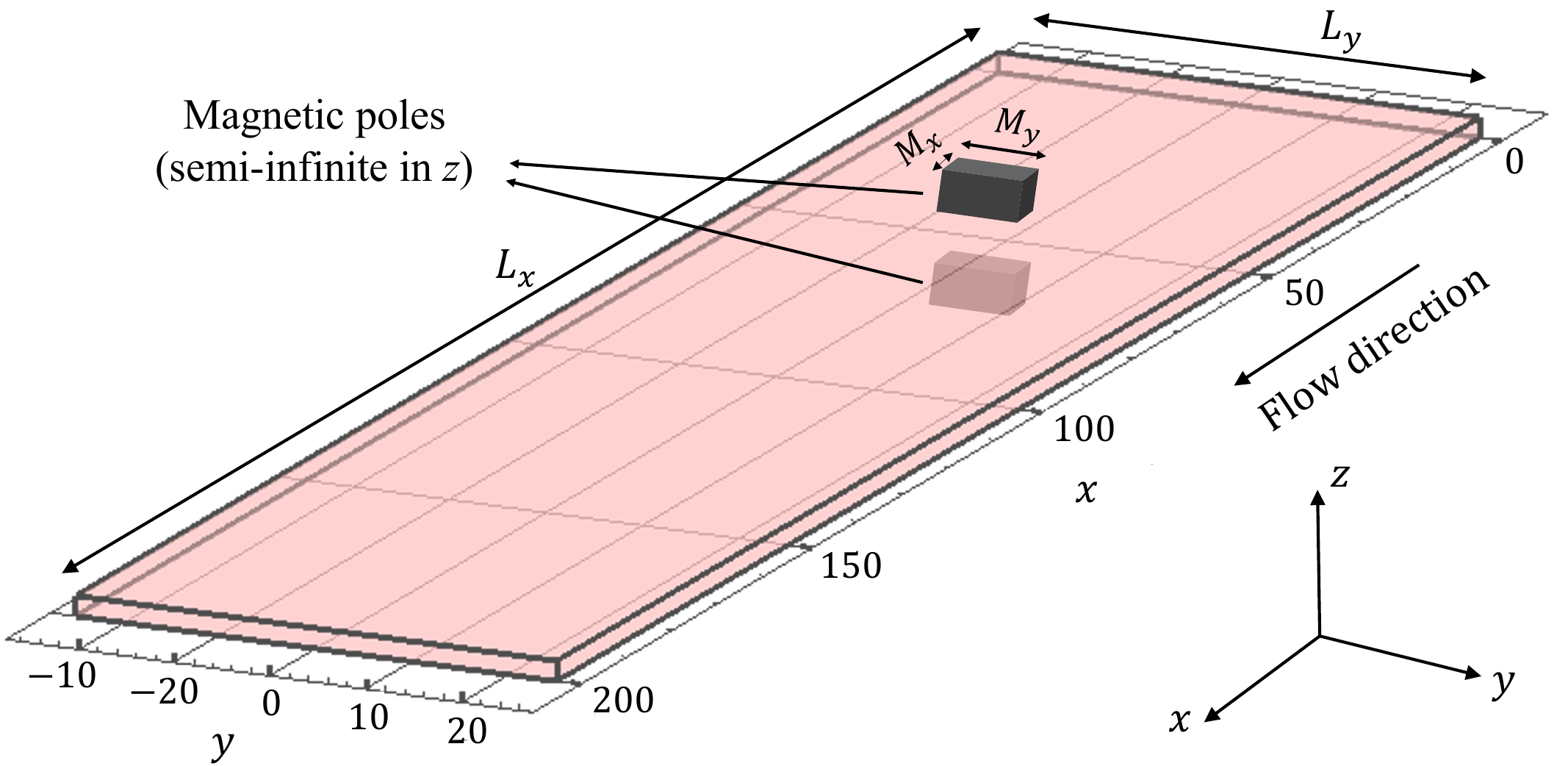}}
	\caption{\textcolor{black}{Schematic of the setup for the duct flow with a localized oscillating magnetic field. The magnetic poles are semi-infinite in $z$-direction and oscillate along $y$-direction. The height $L_z$ of the duct is unity.}}
	\label{fig:Schematic}
\end{figure}

The second configuration, which is the main focus of this paper, consists of liquid metal flow in a duct with two oscillating permanent magnetic poles near the top and bottom walls. The magnetic poles are semi-infinite in the $z$-direction and measure $M_x=3$ units along the streamwise direction and $M_y=4$ units along the spanwise direction in agreement with Votyakov and Kassinos~\cite{Votyakov:PF2009}.  
The spanwise dimension of the duct is $L_y=50$ units and the height is $L_z=1$ unit. The vertical gap between the magnetic poles and the liquid domain (one quarter of the layer height) also corresponds to Ref.\cite{Votyakov:PF2009}.  A schematic diagram of the setup is shown in Fig.~\ref{fig:Schematic}.

The magnetic field $\boldsymbol{B}=(B_x,B_y,B_z)$ generated by the magnetic poles is given by the formula derived by Votyakov \textit{et al.}~\cite{Votyakov:TCFD2009}.
The oscillation takes place along the spanwise direction and the $y$-coordinate of the center of the magnet at time $t$ is given by
\begin{equation}
	y_m=A\sin(2{\pi}f_0t),
\end{equation}
 where $A$ and $f_0$ are the amplitude and frequency of oscillation respectively. The magnets therefore have a  velocity $\boldsymbol{u}_m$ with respect to the flow domain. Since the induction of currents depends on the relative velocity between conductor and magnet, the difference $\boldsymbol{u}-\boldsymbol{u}_m$ must be used in Ohm's law (\ref{eq:Lorentz}b) and in the charge conservation condition (\ref{eq:Lorentz}c).

 In our work, the oscillation amplitude is set to $A=1$, i.e.  the ratio $A/M_y=0.25$. For the frequency we choose a reference value based on the Strouhal number $\St_0=0.25$ 
  in Ref.~\cite{Votyakov:PF2009}.  
 The nondimensional reference frequency in our work is therefore
\begin{equation}
	f_s = \frac{\St_0 U}{M_y} = \frac{0.25 \times 1}{4} = 0.0625
\end{equation}
where $U=1$ is the nondimensional mean streamwise velocity. The frequency ratio is defined as $F=f/f_s$, i.e., the ratio of the frequency of oscillation of the magnetic poles to that of vortex shedding for the stationary magnetic obstacle of the same dimensions at a Reynolds number $\Rey=900$  in Ref.~\cite{Votyakov:PF2009}. 

\section{Numerical method}\label{sec:Numerics}
We conduct direct numerical simulations of our setups using a second-order finite difference code developed by Krasnov \textit{et al.}~\cite{Krasnov:CF2011,Krasnov:JCP2023}.
 For the magnetoconvection setup, a non-uniform grid of resolution $4800 \times 9600 \times 300$ was used.  All the walls were rigid and electrically insulated such that the electric current density $\boldsymbol{j}$ formed closed field lines inside the cell. The top and bottom walls were fixed at $T=-0.5$ and $T=0.5$ respectively, and the sidewalls were adiabatic. The Rayleigh number, Prandtl number, and the Hartmann number based on the maximum value of the vertical magnetic field were fixed at $\Ra=10^5$, $\Pran=0.021$, and $\Ha_{z,max}=120$. The gap $\delta$ between the magnetic poles and the conducting plates was varied from $\delta=0.01H$ to $\delta=9H$, where $H$ is the cell height.

For the configuration of flow past oscillating magnetic obstacle, we employ a rectangular domain of dimensions $L_x \times L_y \times L_z = 200 \times 50 \times 1$ with a grid resolution of $1024 \times 384 \times 32$.   The fluid enters the domain at $x=0$ with a nearly fully-developed laminar flow profile that is approximated by the analytical expression
\begin{equation}
u = \frac{\cosh\left(  \frac{1.55L_y}{L_z}  \left | \frac{2y}{L_y} \right |  \right)-\cosh\left(  \frac{1.55L_y}{L_z}   \right)}{1 - \cosh \left(\frac{1.55L_y}{L_z} \right)} \cdot
\frac{\cosh \left( \frac{1.55L_z}{L_y} \left | \frac{2z}{L_z} \right | \right) - \cosh \left(\frac{1.55L_z}{L_y} \right)}{1-\cosh \left( \frac{1.55L_z}{L_y}\right)}.
\end{equation}
The fluid leaves the domain at $x=L_x$ where $\partial \boldsymbol{u}/\partial x = 0$. The magnetic poles are located $x=50$. The mesh is non-uniform in $y$ and $z$-directions. The top, bottom, and sidewalls are rigid (no-slip) and electrically insulated.  We fix $\Rey=400$ and the Hartmann number based on the maximum vertical magnetic field as $\Ha_{z,max}=70$, and vary the frequency ratio from $F=0.2$ to $F=0.8$. It must be noted that the characteristic length and velocity for the above nondimensional quantities are $L_z/2$ (that is, half of the duct height) and the bulk horizontal velocity at the inlet ($U$), respectively.

For both the simulation setups, the elliptic equations for pressure, electric potential, and the temperature were solved based on applying cosine transforms in along the directions with uniform grid-stretching and using a tridiagonal solver along the direction with non-uniform grid stretching. The diffusive term in the temperature transport equation is treated implicitly. The time discretization of the momentum equation uses the fully explicit Adams-Bashforth/Backward-Differentiation method of second
order.

\section{Results}\label{sec:Results}

In this section, we first briefly summarize the results on the magnetoconvection simulations and then  describe in detail the results on the flow past oscillating magnetic obstacle.
\subsection{Results on magnetoconvection} \label{sec:Result_magnetoconvection}
\begin{figure}
	\centerline{\includegraphics[scale = 0.23]{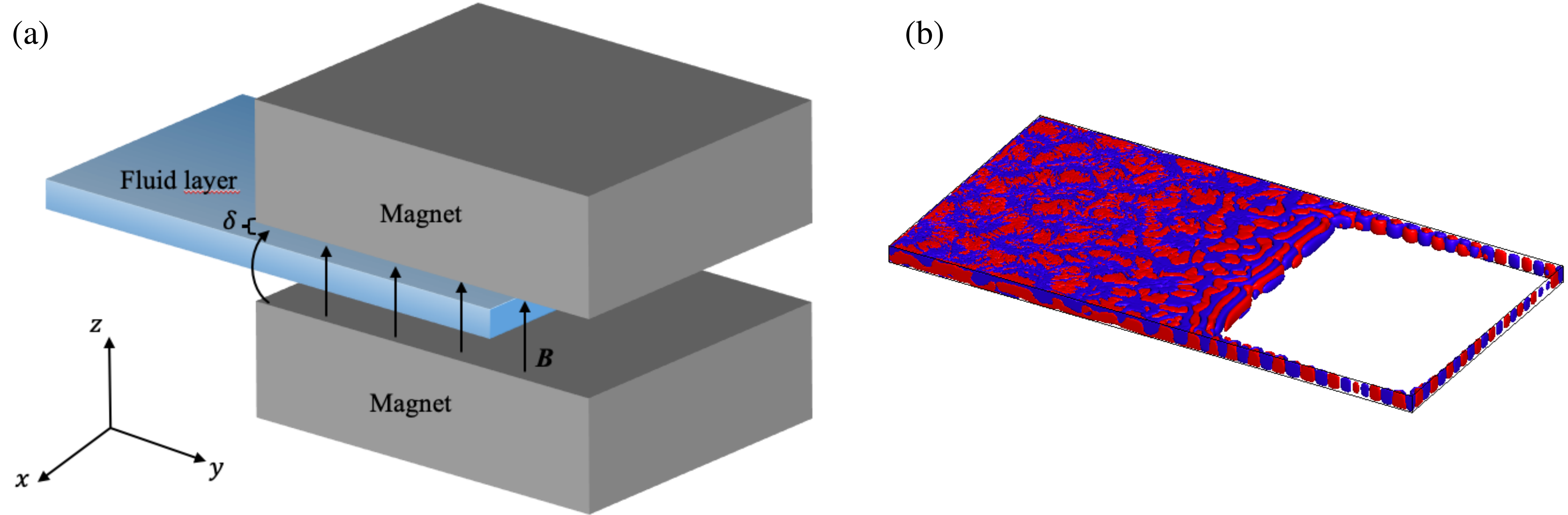}}
	\caption{ 
 (a) Schematic diagram of the magnetoconvection setup, and (b) isosurface contours of vertical velocity $u_z = 0.01$ (red) and $u_z= -0.01$ (blue) for magnetoconvection with $\delta/H=3$ \cite{Bhattacharya:JFM2023}. The magnetic poles are semi-infinite in $y$ and $z$-directions, and infinite along $x$-direction. The  fluid motion in the region of strong magnetic fields is restricted to narrow zones adjacent to the sidewalls.}
	\label{fig:Convection}
\end{figure}
A schematic of the magnetoconvection setup is illustrated in Fig.~\ref{fig:Convection}(a). The magnetic field generated by the magnets is strong enough to cease the flow in  high magnetic flux region of the convection cell. We observe that as the local vertical magnetic field strength increases, the large scale structures become thinner and align themselves perpendicular to the longitudinal sidewalls.
The dependence of the local Reynolds and Nusselt numbers on the local Hartmann number (based on the vertical component of the magnetic field) was determined; this dependence was observed to be independent of the fringe-width.
The global heat transport was observed to decrease with increasing fringe-width for strong magnetic fields but decrease with increasing fringe-width for weak magnetic fields.
The convective motion became confined to the vicinity of the sidewalls in the regions of strong magnetic fields as shown in Fig.~\ref{fig:Convection}(b).
The amplitudes of these wall modes were shown to exhibit a non-monotonic dependence on the fringe-width.

For further details on the results, the readers are referred to Bhattacharya \textit{et al.}~\cite{Bhattacharya:JFM2023}. In the next section, we discuss the results for the MHD duct flow setup.

\subsection{Results on flow past oscillating magnetic obstacle} \label{sec:Result_obstacle}

\begin{figure}
  	\centerline{\includegraphics[scale = 0.22]{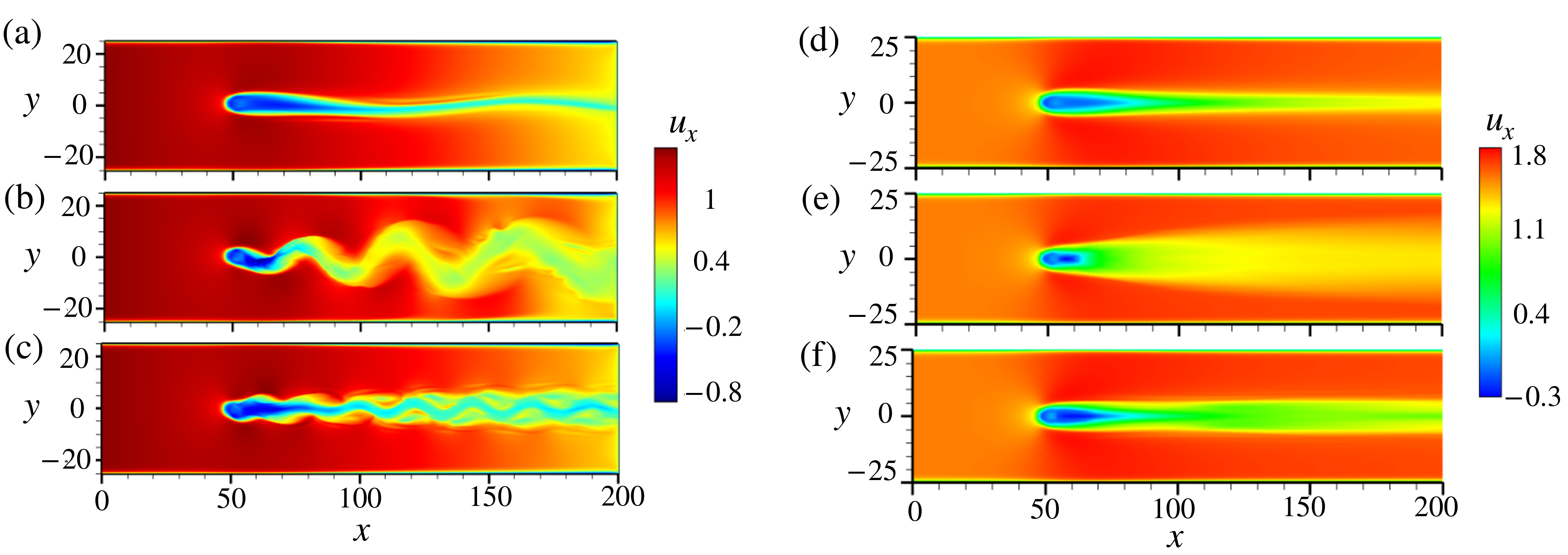}}
  	\caption{Contour plots of streamwise velocity $u_x$ for flows past oscillating magnetic obstacle for different frequency ratios in the midplane $z=0$. Instantaneous contour plots for (a) $F=0.2$, (b) $F=0.5$, and (c) $F=0.8$. Time-averaged contour plots for (d) $F=0.2$, (e) $F=0.5$, and (f) $F=0.8$.}
  	\label{fig:Vel_density}
  \end{figure}

 The simulations of the flow past magnetic obstacles are run for 300 convective time units after reaching a fully-developed state. The contour plots of instantaneous streamwise velocity are exhibited in Figs.~\ref{fig:Vel_density}(a--c) and those with  time-averaging in Figs.~\ref{fig:Vel_density}(d--f) for $F=0.2$, $F=0.5$, and $F=0.8$. The figures show regions of reduced and even reversed streamwise velocity in the regions of strong magnetic field and also in the wake of the magnetic obstacle. The regions of reduced instantaneous velocity exhibit a wavy pattern. It can be visually observed from Figs.~\ref{fig:Vel_density}(a--c) that as the magnets oscillate faster, the wavelength of spatial oscillation decreases. There is an increase in the amplitude of this path from $F=0.2$ to $F=0.5$, but the amplitude decreases with a further rise in $F$. For $F=0.5$, the wake of the magnetic obstacle comprises of small-scale eddies, indicating that the flow becomes turbulent. The time-averaged streamwise velocity contours show that the length of the reversed flow region first decreases as $F$ is increased to 0.5, and then increases with a further increase of $F$. 

\begin{figure}
 	\centerline{\includegraphics[scale = 0.22]{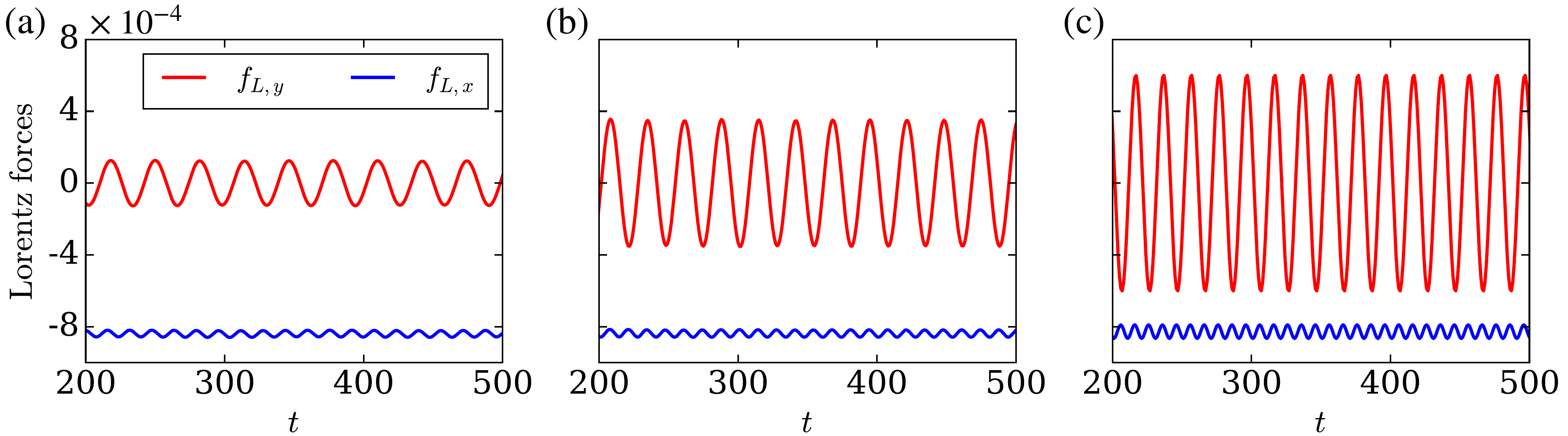}}
 	\caption{Time-series of the total spanwise and streamwise component of Lorentz force ($f_{L,y}$ and $f_{L,x}$, respectively) in flow past oscillating magnetic obstacles for the following frequency ratios: (a) $F=0.5$, (b) $F=0.6$, and (c) $F=0.8$. The Lorentz force exhibits a sinusoidal time-dependence.}
 	\label{fig:Lift_drag}
 \end{figure}

\begin{figure}
 	\centerline{\includegraphics[scale = 0.18]{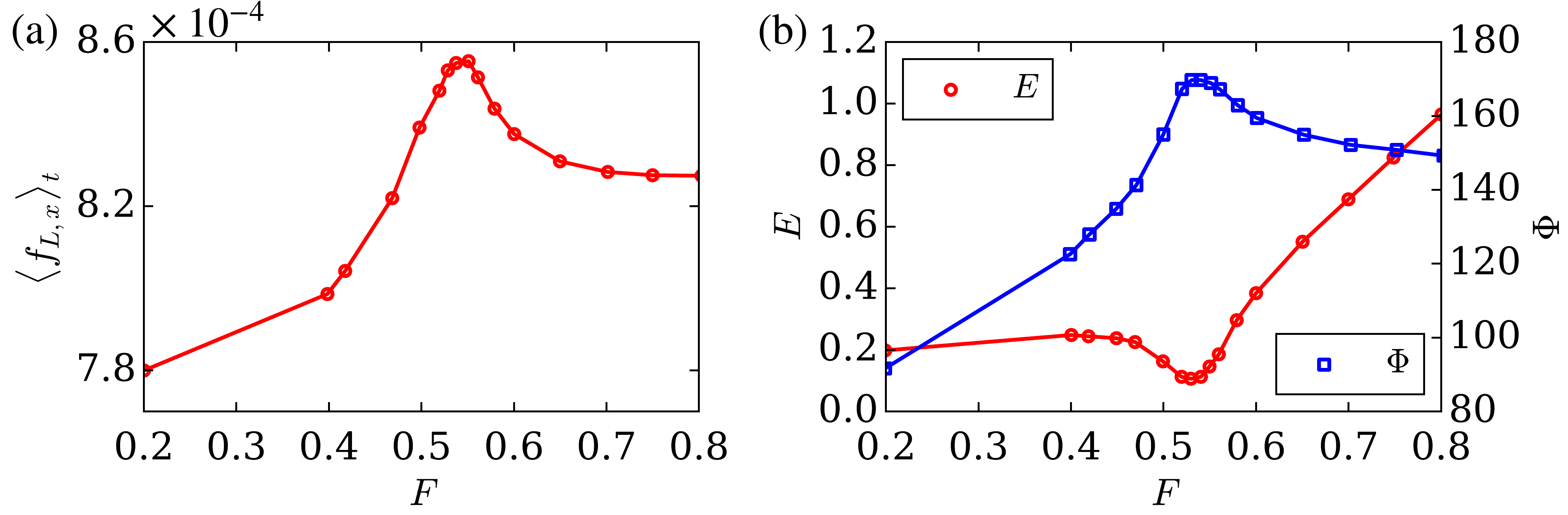}}
 	\caption{For flow past oscillating magnetic obstacle: (a) Variation of the time-averaged Lorentz force in the streamwise direction with the frequency ratio, and (b) plots of energy transferred from the magnets to the fluid and the phase angle between the spanwise Lorentz force and displacement of the magnets. The above quantities exhibit a nonmonotonic dependence on $F$. }
 	\label{fig:Energy}
 \end{figure}

 We examine the components of the total Lorentz force  in the streamwise ($f_{L,x}$) and spanwise ($f_{L,y}$) directions. Note that $f_{L,x}$ and $f_{L,y}$ are the analog of the drag and lift forces, respectively, in flow past solid cylinders. These quantites are
 \begin{equation}
 	f_{L,x} = \int_{-L_z/2}^{L_z/2} \int_{-L_y/2}^{L_y/2} \int_0^{L_x} (\boldsymbol{f}_{L} \cdot \boldsymbol{\hat{x}})\, dx \,dy \,dz,
 	\quad
 	f_{L,y} = \int_{-L_z/2}^{L_z/2} \int_{-L_y/2}^{L_y/2} \int_0^{L_x} (\boldsymbol{f}_{L} \cdot \boldsymbol{\hat{y}})\, dx\, dy \,dz.
 \end{equation}
 Figures~\ref{fig:Lift_drag}(a,b,c) exhibit the plots of the above quantities versus the convective time $t$ for $F=0.5$, $F=0.6$, and $F=0.8$, respectively. The figures show a periodic sinusoidal time dependence. The magnitude of $f_{L,x}$ is higher than $f_{L,y}$; however, $f_{L,y}$ oscillates with a higher amplitude than $f_{L,x}$. The amplitude of oscillation increases with an increase of $F$. The response frequency of $f_{L,y}$ is equal to the excitation frequency $f_0$ of the magnets; however, the response frequency of $f_{L,x}$ is twice of $f_0$.
 We further compute $\langle f_{L,x} \rangle_t$, the streamwise component of Lorentz force averaged over 300 timeframes, and plot it versus the frequency ratio in Fig.~\ref{fig:Energy}(a). It can be seen that $\langle f_{L,x} \rangle_t$ increases rapidly from $F=0.2$ to $F=0.55$. On further increase of $F$,  $\langle f_{L,x} \rangle_t$ decreases sharply till $F=0.7$ above which $\langle f_{L,x} \rangle_t$ saturates close to a constant value.
 Interestingly, the aforementioned behaviors of $f_{L,x}$ and $f_{L,y}$ closely resemble that of the drag and lift forces, respectively, in flows past an oscillating cylinder~\cite{Blackburn:JFM1999,Placzek:JCF2009}.

For flows past an oscillating cylinder, the non-dimensional mechanical energy transferred from the cylinder to the fluid is expressed as
 \begin{equation}
 	E = \frac{2}{\rho U d^2}\int_{0}^{t_P}\frac{dy}{dt}f_\mathrm{lift}\, dt,
 	\label{eq:Energy_cylinder}
 \end{equation}
where $t_P$ is the motion period, $d$ is the diameter of the cylinder, $y$ is the spanwise position of the cylinder's axis, and $f_\mathrm{lift}$ is the magnitude of the lift force~\cite{Blackburn:JFM1999}. In our work, the energy transferred from the  oscillating magnets to the fluid can be expressed similarly as follows:
 \begin{equation}
	E = \int_{0}^{t_P}\frac{dy_m}{dt}\, f_{L,y}\,dt,
	\label{eq:Energy_magnet}
\end{equation}
where $t_P=$ 300 convective time units for our case.
We compute $E$ for different frequency ratios and plot it versus $F$ in Fig.~\ref{fig:Energy}(b). The figure shows that $E$ is always positive, implying that the magnets perform work on the fluid for all frequencies. The figure further shows that there is a gradual growth of $E$ until $F=0.45$ and then it sharply decreases to a minimum value at $F=0.53$. The energy transfer increases monotonically on further increase of $F$.  Interestingly, the point of minimum energy transfer almost coincides with the point of maximum average streamwise Lorentz force. This point corresponds to resonance where the velocity field exhibits stronger fluctuations compared to other frequency ratios.

 We finally examine the trends of the phase angle between the spanwise component of Lorentz force and the spanwise displacement of the magnets.
 This parameter is used as an indicator of energy transfer from the magnets to the fluids~\cite{Blackburn:JFM1999} where a phase angle between 0 and 180 degrees indicates positive energy transfer.
 We compute the phase angle using our data by fitting it with a sinusoidal function using the method of least squares.
  The computed phase angle is plotted versus the frequency ratio in Fig.~\ref{fig:Energy}(b). The figure shows that the phase angle lies between 0 and 180 degrees, consistent with the fact that the energy is transferred from the magnets to the fluid. The maximum phase angle at $F=0.55$ reaches about 170 degrees. 
  


  \section{Conclusions}\label{sec:Conclusions}
    In this paper, we numerically examined the effects of non-homogeneous magnetic fields in liquid metal flows using a finite-difference fluid solver. We briefly summarized the results of Bhattacharya \textit{et al.}~\cite{Bhattacharya:JFM2023} in which the influence of fringing magnetic fields on turbulent convection was studied. An important finding was that for strong magnetic fields, the global heat transport decreases with an increase of fringe-width, whereas for weak magnetic fields, the heat transport marginally increases with an increase of fringe-width. The convective motion gets confined near the sidewalls in regions of strong magnetic fields.

    We numerically examined the effects of oscillating magnetic obstacle with different frequencies on liquid metal flow in a duct.
    We showed the presence of reduced and reversed streamwise velocity in the regions of strong magnetic field and in the wake of the magnetic obstacle. The regions of reduced velocity exhibit a wavy pattern with the wavelength of spatial oscillation decreasing with the excitation frequency of the magnets. The amplitude of wake oscillation shows a non-monotonic dependence on the frequency of the magnets and exhibits a maximum at a particular frequency $f_{\text{max}}$, which  appears to correspond to the point of maximum Lorentz force in the streamwise direction and the minimum work done by the magnets on the fluid. The total streamwise and spanwise components of the Lorentz force oscillate sinusoidally with time with the same frequency and twice the frequency of the magnet's oscillation. The mean of the spanwise Lorentz force is zero. Its amplitude  increases with an increase of the frequency of oscillation of the magnets. The frequency $f_{\text{max}}\approx 0.5 f_s$ is considerably smaller than the reference value $f_s$ taken from Ref.~\cite{Votyakov:PF2009}. Although the stationary magnet does not produce vortex shedding in our case, it seems plausible that our value $f_{\text{max}}$ is indicative of the intrinsic shedding frequency when $\Rey$ (and possibly $\Ha$)  are increased further. Lower values than $\St_0=0.25$ of the Strouhal number of stationary magnetic obstacles were also reported by Kenjere\v{s}  \textit{et al.}~\cite{Kenjeres:IJHFF2011}.

%
%
%
%

\begin{acknowledgement}
 The authors are grateful to J. Schumacher for providing valuable contributions to the study of convection under the influence of fringing magnetic fields. S. Bhattacharya is supported by a postdoctoral fellowship of Alexander von Humboldt Foundation, Germany.
\end{acknowledgement}

\vspace{\baselineskip}

\bibliographystyle{pamm}
\providecommand{\WileyBibTextsc}{}
\let\textsc\WileyBibTextsc
\providecommand{\othercit}{}
\providecommand{\jr}[1]{#1}
\providecommand{\etal}{~et~al.}

\end{document}